\documentclass[aps,pra,twocolumn,amsmath,amssymb,nofootinbib,superscriptaddress]{revtex4}

\newcommand{\ket}[1]{|#1\rangle}

\usepackage[pdftex]{graphicx}
\usepackage{mathrsfs}
\usepackage{comment}
\usepackage[colorlinks]{hyperref}

\begin{document}

\bibliographystyle{apsrev}

\title{Scalable boson-sampling with time-bin encoding using a loop-based architecture}

\author{Keith R. Motes}
\affiliation{Centre for Engineered Quantum Systems, Department of Physics and Astronomy, Macquarie University, Sydney NSW 2113, Australia}

\author{Alexei Gilchrist}
\affiliation{Centre for Engineered Quantum Systems, Department of Physics and Astronomy, Macquarie University, Sydney NSW 2113, Australia}

\author{Jonathan P. Dowling}
\affiliation{Hearne Institute for Theoretical Physics and Department of Physics \& Astronomy, Louisiana State University, Baton Rouge, LA 70803}
\affiliation{Computational Science Research Center, Beijing 100084, China}

\author{Peter P. Rohde}
\email[]{dr.rohde@gmail.com}
\homepage{http://www.peterrohde.org}
\affiliation{Centre for Engineered Quantum Systems, Department of Physics and Astronomy, Macquarie University, Sydney NSW 2113, Australia}

\date{\today}

\frenchspacing

\begin{abstract}
We present an architecture for arbitrarily scalable boson-sampling using two nested fiber loops. The architecture has fixed experimental complexity, irrespective of the size of the desired interferometer, whose scale is limited only by fiber and switch loss rates. The architecture employs time-bin encoding, whereby the incident photons form a pulse train, which enters the loops. Dynamically controlled loop coupling ratios allow the construction of the arbitrary linear optics interferometers required for boson-sampling. The architecture employs only a single point of interference and may thus be easier to stabilize than other approaches. The scheme has polynomial complexity and could be realized using demonstrated present-day technologies.
\end{abstract}

\maketitle

Quantum computing is of major interest today as it promises exponentially improved performance in some key algorithms \cite{bib:NielsenChuang00}. One of the more promising schemes for realizing a quantum computer is linear optics quantum computing (LOQC) \cite{bib:KokLovett11}, first presented by Knill, Laflamme \& Milburn (KLM) \cite{bib:KLM01}. This seminal result shows that universal quantum computing is possible using only single-photon sources, linear optics elements, and photodetectors. However, full-fledged LOQC has daunting physical resource requirements, perhaps requiring millions of optical elements to build a post-classical device. These massive requirements have stimulated interest in finding simpler, albeit non-universal, approaches to LOQC, requiring fewer physical resources.

One such proposal is boson-sampling, first presented by Aaronson \& Arkhipov \cite{bib:AaronsonArkhipov10}, whereby a passive linear optics network is fed with single photon sources and measured via coincidence photodetection. It has been estimated that building a post-classical device via this approach might only require dozens of photons and hundreds of optical elements, a significant improvement over KLM. This approach to LOQC is not universal for quantum computation, but is strongly believed to implement a classically inefficient algorithm. Several elementary experimental demonstrations of boson-sampling have recently been performed \cite{bib:Peruzzo17092010, bib:Broome2012, bib:Tillmann4, bib:Crespi3, bib:Spring2}.

In the boson-sampling model we begin by preparing $p$ single photons in $n$ modes,
\begin{equation}
\ket{\psi}_\mathrm{in} = \ket{1_1,\dots,1_p,0_{p+1},\dots,0_n},
\end{equation}
where \mbox{$n=O(p^2)$}.

This input state propagates through a passive linear optics network, which implements a unitary map $\hat{U}$ on the photon creation operators,
\begin{equation}
\hat{U}: \,\, \hat{a}_i^\dag \to \sum_j U_{i,j} \hat{a}^\dag_j,
\end{equation}
where $\hat{a}^\dag_i$ is the photon creation operator in the $i$th mode.

The output state is of the form,
\begin{equation}
\ket{\psi}_\mathrm{out} = \sum_S \gamma_S \ket{p_1^{(S)},\dots,p_n^{(S)}},
\end{equation}
where $S$ are the different photon-number configurations, and \mbox{$p_i^{(S)}$} is the number of photons in the $i$th mode associated with configuration $S$. The number of terms in the superposition scales as \mbox{$|S|=\binom{n+p-1}{p}$}, which grows exponentially with $p$. The total photon number is conserved. Thus, \mbox{$\sum_i p_i^{(S)} = p \,\,\forall \,S$}, and upon measurement we always detect $p$ photons in total.

Finally we measure the output state using coincidence number-resolved photodetection, and each time we repeat the experiment we collapse onto a particular term, thereby sampling the photon-number superposition. This sampling problem was shown by Aaronson \& Arkhipov to likely be not classically simulatable. While the full complexity proof is complex, the intuitive explanation for the computational hardness of boson-sampling is that: (1) there are an exponential number of terms in the output superposition, which rules out brute force simulation of the state vector; and (2) the amplitude associated with each term in the output superposition is related to a matrix permanent \cite{bib:Scheel04}, whose calculation resides in the complexity class \textbf{\#P}-complete, a class strongly believed to be classically hard to simulate. Specifically, \mbox{$\gamma_S \propto \mathrm{Per}(U_S)$}, where \mbox{$U_S$} is a \mbox{$p\times p$} submatrix of $U$ as a function of the respective output configuration $S$. Note that each amplitude is proportional to a different permanent, which rules out using boson-sampling to calculate matrix permanents, as this would require an exponential number of measurements. 

\begin{figure}[!htb]
\includegraphics[width=0.5\columnwidth]{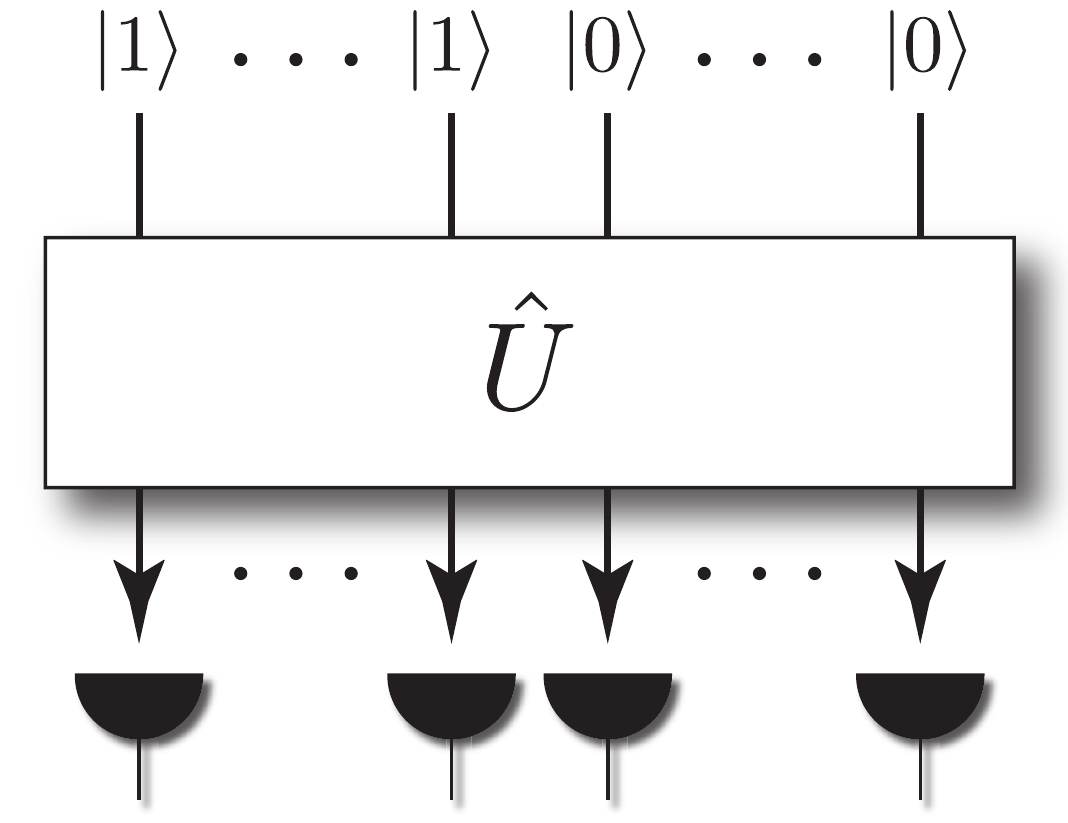}
\caption{The boson-sampling model. A series of single photon and vacuum states are prepared and propagated through a linear optics network. The output state is sampled via coincidence photodetection.} \label{fig:model}
\end{figure}

The full model is illustrated in Fig.~\ref{fig:model}.

Presently, various technologies are available for preparing single photon states (e.g spontaneous parametric down-conversion, which has been shown to be viable for boson-sampling \cite{bib:Motes13, bib:LundScatter13}), and performing photodetection (e.g avalanche photodiodes). The remaining central challenge in boson-sampling is constructing linear optics networks $\hat{U}$. It was shown by Reck \emph{et al.} \cite{bib:Reck94} that arbitrary networks of this form can be decomposed into a sequence of \mbox{$O(n^2)$} beamsplitters. In present-day experiments this type of decomposition is implemented using waveguides or discrete optical elements. 

In current implementations, the modes in a boson-sampling interferometer are spatial modes, whereby all photons must have simultaneous arrival time. Another alternative is to employ time-bin encoding, whereby $n$ single photons form a `pulse train' within a single spatial mode. In the architecture we present here, we will employ time-bin encoding.

\begin{figure}[!htb]
\includegraphics[width=0.8\columnwidth]{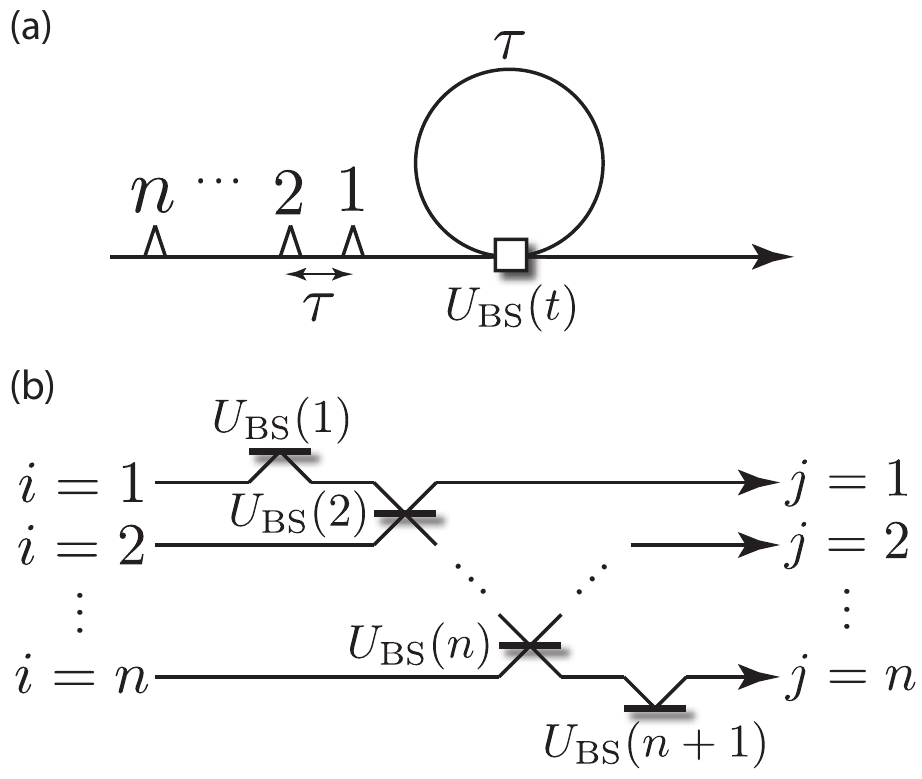} 
\caption{(a) A fiber loop fed by a pulse train of single photons, each separated in time by $\tau$ (or length $\tau$ in units of \mbox{$c=1$}). The box represents a dynamically controlled, variable reflectivity beamsplitter, given by the beamsplitter matrix $U_{\mathrm{BS}}(t)$ at time $t$. The switching time of the beamsplitter must be less than $\tau$ to allow each time-bin to be individually addressed. (b) Expansion of the fiber loop architecture into its equivalent beamsplitter network, obtained by `unravelling' the loop and mapping time-bins to spatial modes.} \label{fig:singleloop}
\end{figure}

We begin by triggering a single photon source at time intervals $\tau$ (the source's repetition rate), which prepares a pulse train of $n$ single photons across a length of fiber. The first step in our architecture is to propagate the pulse train through a fiber loop with dynamically controlled coupling ratios, as shown in Fig.~\ref{fig:singleloop}(a). The loop's coupling ratio is dynamically controlled by a variable reflectivity beamsplitter, implementing the unitary,
\begin{equation} \label{eq:VarBS}
U_{\mathrm{BS}}(t) = \left[ \begin{array}{cc}
\gamma_{1,1}(t) & \gamma_{1,2}(t)  \\
\gamma_{2,1}(t) & \gamma_{2,2}(t) 
\end{array} \right], 
\end{equation}
at time $t$, which splits the incident field into a component entering the loop and a component exiting the loop. The component entering the loop takes time $\tau$ to transverse the loop such that it coincides with the subsequent pulse. In order for the first photon to interfere with every photon pulse it will traverse this loop \mbox{$n$} times. The second photon will traverse the loop \mbox{$n-1$} times and so on. The dynamics of photons propagating through the loop architecture may be `unravelled' into an equivalent series of beamsplitters acting on spatial modes, as shown in Fig.~\ref{fig:singleloop}(b). This elementary network is the basic building block employed by our architecture.

The effective unitary map applied to the time-bins through a single such loop is,
\begin{equation}
U_{i,j} = \left\{ \begin{array}{ll}
 0 & i>j+1 \\
 \gamma_{2,1}(i) & i=j+1 \\
 \gamma_{2,2}(i)\gamma_{1,1}(j+1) \prod_{k=i+1}^{j}{\gamma_{1,2}}(k) & i<j+1
 \end{array} \right. ,
\end{equation}
where $i$ and $j$ represent input and output modes respectively. Here we have imposed the boundary condition that $U_{\mathrm{BS}}(1)$ and $U_{\mathrm{BS}}(n+1)$ are completely reflective, coupling all of the first photon into the loop, and ensuring that all of the field remains trapped in a finite time-window. We see that the probability of finding a photon in the $j$th mode decays exponentially with $j$. 

Evidently, the network shown in Fig.~\ref{fig:singleloop}(b) is not sufficient for universal linear optics networks as it contains many zero elements (note that by `universal linear optics network' we mean that arbitrary linear optics networks may be constructed, which is distinct from universal quantum computing). To make the scheme universal we must show that the ingredients necessary to perform a full Reck \emph{et al.}-type decomposition are available (such decompositions find many other uses beyond boson-sampling).

\begin{figure}[!htb]
\includegraphics[width=.85\columnwidth]{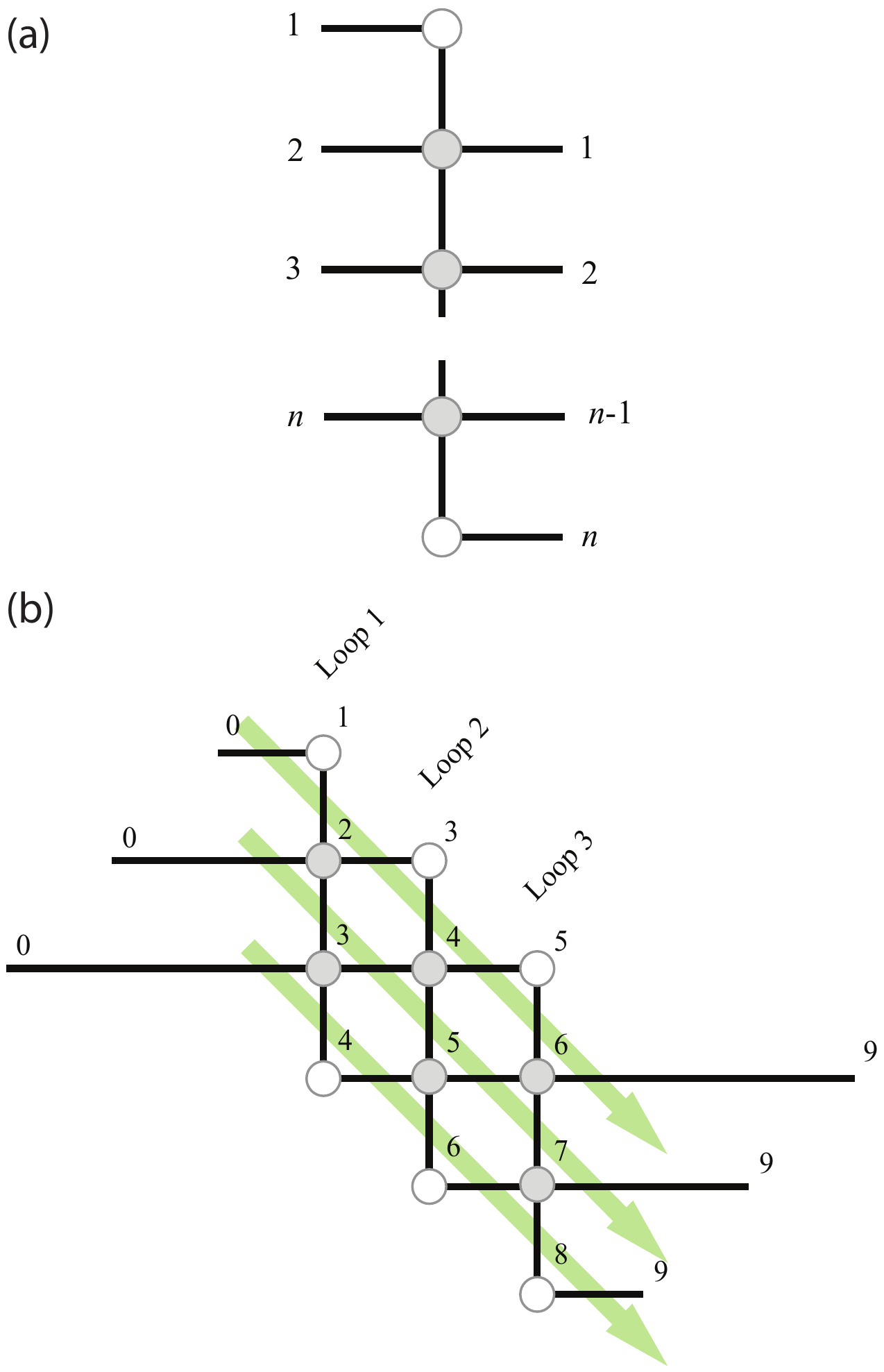} 
\caption{(Color online) The equivalent beam-splitter representation for the fiber loop architecture. Closed circles represent beamsplitter operations, open circles are fully reflective operations, and the numbers represent the time-bin associated with a path. (a) A single loop is represented for $n$ photons in the pulse train. The first photon is deterministically coupled into the loop and interferes with the second photon. At this point some of the amplitude leaves the loop which represents the first time-bin. This process in continued until the $n$th photon transverses the loop, whereby any remaining amplitude is deterministically coupled out. (b) The equivalent beamsplitter network of three consecutive loops with three input modes (horizontal lines, top-left). The lengths of the black lines represent time in units of $\tau$. The three modes on the left represent the pulse train of photons at the input of the device at the first round-trip. The first photon reaches the first beamsplitter at \mbox{$\tau=1$}, the second photon reaches it at \mbox{$\tau=2$}, and so on. The photons travel through the fiber loop network interacting arbitrarily, which yields an arbitrary Reck \emph{et al.}-style decomposition.} \label{fig:AlexeiProof}
\end{figure}

To understand the equivalent beamsplitter representation of a single loop, consider Fig.~\ref{fig:AlexeiProof}(a). The pulse train enters the loop, where the numbers on the left represent the corresponding time-bin. The first photon is deterministically coupled into the loop as depicted by an open circle. After the first and second photons interact some of the amplitude may escape the loop, which corresponds to the first output time-bin. The pulse train continues to interact through the loop via beamsplitter operations, which are represented as closed circles. After the $n$th photon transverses the loop any remaining amplitude deterministically leaves the loop, which corresponds to the $n$th output time-bin.

Now consider Fig.~\ref{fig:AlexeiProof}(b), which depicts how three consecutive loops in series with three input photons produce an equivalent beamsplitter network. Evidently, with three loops and three input photons arbitrary pairwise beamsplitter interactions can be implemented allowing for the implementation of an arbitrary unitary. This easily extends to implementing an $n$-mode unitary by adding additional modes and loops. An alternate inductive proof is shown in App.~\ref{app:alt_proof}.

\begin{figure}[!htb]
\includegraphics[width=0.8\columnwidth]{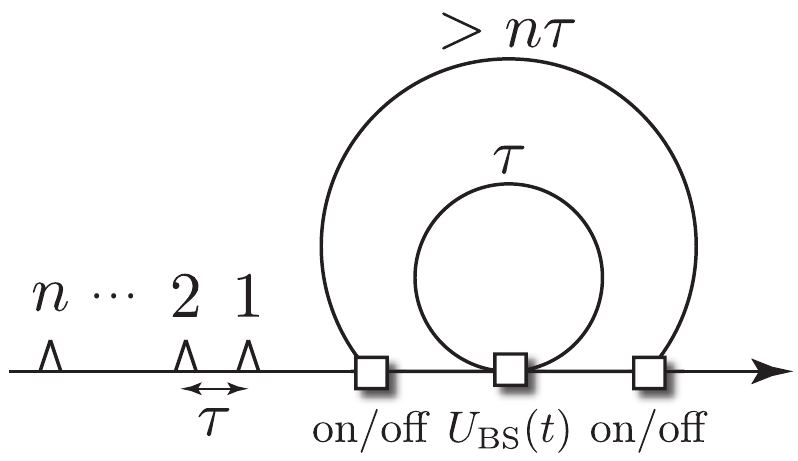} 
\caption{The complete architecture. The consecutive series of length $\tau$ fiber loops is collapsed into a single length $\tau$ fiber loop embedded inside a length \mbox{$>n\tau$} fiber loop. The outer loop allows an arbitrary number of the smaller loops to be applied consecutively.} \label{fig:full_architecture}
\end{figure}

We have shown that a series of consecutive fiber loops can implement an arbitrary sequence of pairwise beamsplitter operations. Next, we recognize that each of these fiber loops requires exactly the same physical resources, only differing by the switch's control sequence. We need not physically build each of these identical loops. Rather, we will embed the loop into a larger fiber loop of length \mbox{$>n\tau$}, as shown in Fig.~\ref{fig:full_architecture}. The larger loop is controlled by another two switches, which control the number of round trips in the larger loop. From the result of Reck \emph{et al.} we know that \mbox{$O(n^2)$} optical elements are required to construct an arbitrary \mbox{$n\times n$} interferometer. Thus, the number of round trips of the outer loop is $O(n^2)$. 

An experimental simplification is when we do not require full dynamic control over the beamsplitter ratio. Although this scenario is not universal, it may be possible to construct useful classes of unitaries. We will consider the situation where the beampslitter can be toggled between two settings -- completely reflective, or some other arbitrary fixed ratio. The former is required to allow that the time-bins be restricted to a finite time-window, whilst the latter implements the `useful' beamsplitter operations. We may have an arbitrary number of such loops in series, each with a potentially different fixed beamsplitter ratio.

Intuitively, we expect that a `maximally mixing' unitary (i.e. one with equal amplitudes between every input/output pair) would implement a classically hard boson-sampling instance, as it maximizes the combinatorics associated with calculating output amplitudes. Recall that it is the combinatorial complexity of calculating the permanent that gives rise to the computational hardness of boson-sampling. If, for example, a unitary is heavily biased towards certain output modes, or is sparse, the combinatorics are reduced. Specifically, we define a balanced unitary as,
\begin{equation}
|U'_{i,j}|^2=\frac{1}{n} \,\,\forall \,\,i,j,
\end{equation}
such that, up to phase, all amplitudes are equal. Additionally, balanced unitaries, which includes the Hadamard transform and discrete Fourier transform, find many uses in quantum algorithms.

\begin{figure}[!htb]
\includegraphics[width=0.9\columnwidth]{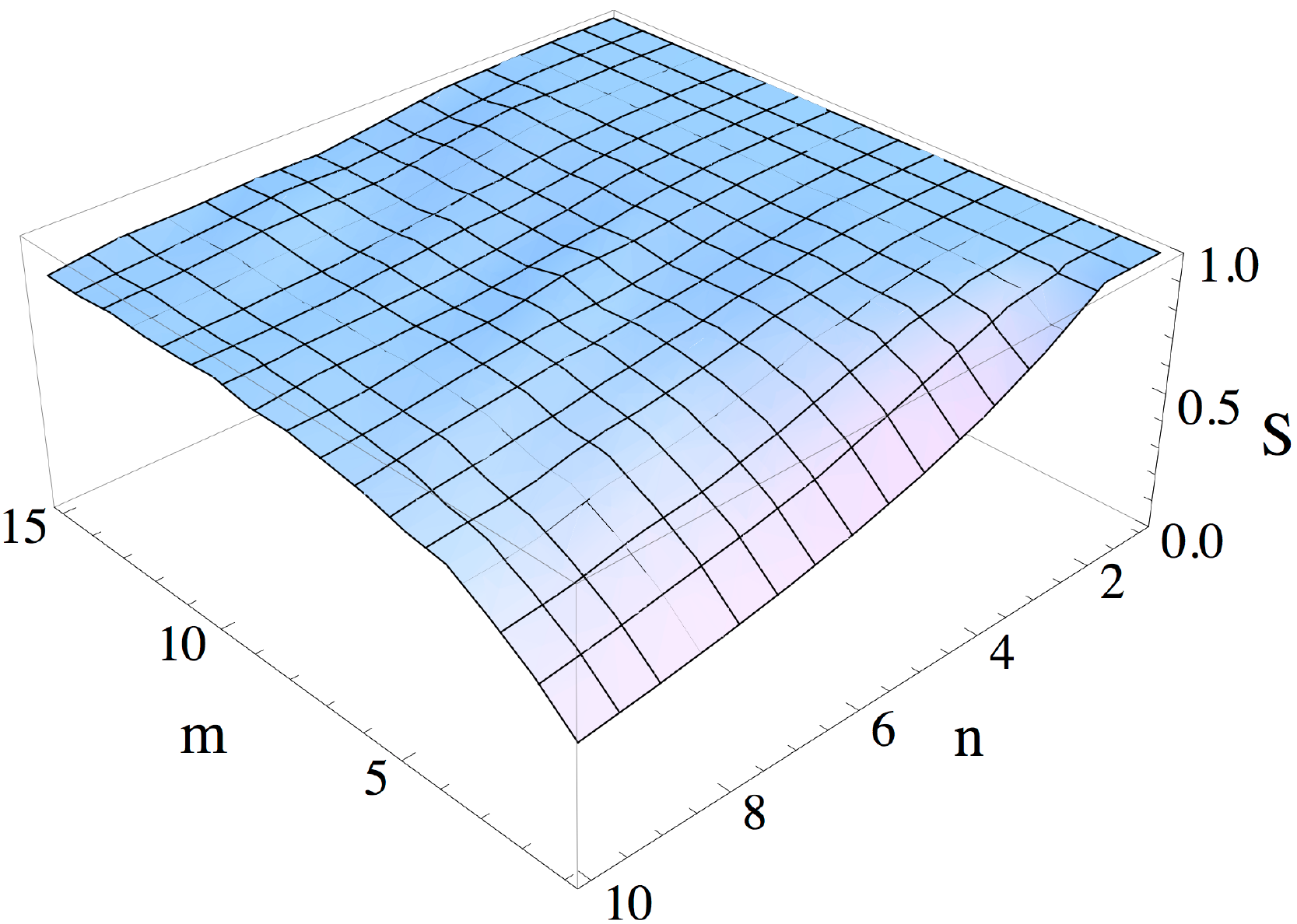} 
\caption{The maximum similarity $S$ between $\hat{U}$ and the uniform unitary $\hat{U}'$ after $m$ loops with $n$ input photons. The beam-splitter ratio is fixed for each loop but independently, randomly chosen for each loop. The first and last beamsplitters in each loop are completely reflective to constrain the photons within the first $n$ time-bins. $S$ tends to decrease with increasing $n$ and increase with increasing $m$, demonstrating that near-uniform unitaries may be constructed with sufficient loops.} \label{fig:similarity}
\end{figure}

In Fig.~\ref{fig:similarity} we take the unitary implemented by a series of $m$ fixed-ratio fiber loops (except the first and last beamsplitters of each loop, which are completely reflective), and compare it with the balanced unitary $\hat{U}'$. We characterize the uniformity of the obtained unitary using the similarity metric,
\begin{eqnarray}
S &=& \max_{\hat{U}_\mathrm{BS}(t) \,\forall\, t}\left[\frac{\left(\sum_{i,j}\sqrt{|U_{i,j}|^2 \cdot |U'_{i,j}|^2}\right)^2}{\left(\sum_{i,j}|U_{i,j}|^2\right) \cdot \left(\sum_{i,j}|U'_{i,j}|^2\right)}\right] \\ \nonumber
&=& \max_{\hat{U}_\mathrm{BS}(t)\,\forall\, t}\left[\frac{1}{n^3}\left(\sum_{i,j}|U_{i,j}|\right)^2\right],
\end{eqnarray}
where we maximize $S$ by performing a Monte-Carlo simulation over different beamsplitter ratios, $\hat{U}_\mathrm{BS}$. That is, $S$ tells us how close $\hat{U}$ is to uniform, with \mbox{$S=1$} being completely uniform up to phase. We see that with a sufficient number of loops in series, we obtain very high similarities, suggesting that the simplified architecture may be useful for boson-sampling, and hard boson-sampling instances might be implemented.

The presented universal architecture is in principle arbitrarily scalable, provided the length of the larger loop is sufficiently high (\mbox{$>n\tau$}). However, in practice, fiber is very lossy with present-day technology. If we let $\eta_\mathrm{inner}$ be the net efficiency of the inner loop (i.e. the probability than an incident photon will reach the ouput), and $\eta_\mathrm{outer}$ be the net efficiency of the outer loop, then the worst case net efficiency of the device is,
\begin{equation}
\eta_\mathrm{net} = ({\eta_\mathrm{inner}}^n \eta_\mathrm{outer})^{O(n^2)},
\end{equation}
which scales exponentially with $n$. Thus, to construct large interferometers using this architecture will require exponentially small loss rates in the fiber loops. However, whilst this may appear drastic, this is also the case for conventional spatially encoded implementations. 
However, it was shown by Rohde \& Ralph \cite{bib:RohdeRalphAA12} that boson-sampling might remain a computationally hard problem even in the presence of very high loss rates. Other error models, such as dephasing or mode-mismatch \cite{bib:RohdeLowFid12}, exhibit similar scaling characteristics.

Because there is only a single point of interference in this architecture (the dynamic beamsplitter controlling the inner loop), this architecture may be significantly easier to stabilize and mode-match than conventional approaches, where \mbox{$O(n^2)$} independent beamsplitters must be simultaneously aligned and stabilized. At this point of interference, the dominant source of error will be temporal mode-mismatch \cite{bib:RohdeRalph05}, which is caused by errors in the lengths of the fiber loops, or time-jitter in the photon sources. Temporal mismatch may be regarded as a displacement in the temporal wavepacket of the photons \cite{bib:RohdeMauererSilberhorn07}. Let us assume that at each round trip the photon exiting the inner loop is mismatched by time $\Delta$. Over short time scales this yields dephasing \cite{bib:RohdeRalph06}, and over longer time scales, ambiguity as to which time-bin the photon resides in. The worst case is that a given photon undergoes temporal mismatch of magnitude $n\Delta$. Time-bin ambiguity occurs when \mbox{$n\Delta \geq \tau$}, which yields the requirement that \mbox{$n<\tau/\Delta$}. Over shorter timescales, temporal mode-mismatch is equivalent to dephasing as mismatched photons yield which-path information. This leads to the constraint that \mbox{$n\Delta \ll \sigma$}, where $\sigma$ is the width of the photons' wave packets. Thus, time-jitter or temporal mode-mismatch must be kept small relative to the scale of the photons' wavepackets.

In principle, the fiber loops could be replaced by any quantum memory (or delay line) such as propagation in free-space (depending on $\tau$), which would be significantly less lossy. In this case, the dominant source of loss would be in the dynamic switches, which, using present-day technology, have very high loss rates.

The experimental viability of loop-based photonic architectures was validated by recent quantum walk \cite{bib:ADZ} experiments by Schreiber \emph{et al.} \cite{bib:Schreiber12, bib:Schreiber10}, where quantum memories were implemented via delay lines in free-space. It was also shown by Donohue \emph{et al.} \cite{bib:Resch} that transmitting time-bin encoded photons in optical fibers is a robust form of optical quantum information given that the separation of time-bins is larger than the time resolution of the detector.

We have presented an arbitrarily scalable architecture for universal boson-sampling based on two nested fiber loops. The complexity of the architecture is constant, independent of the size of the interferometer being implemented. Scalability is limited only by fiber and switch transmission efficiencies. There is only one point of interference in the architecture, which suggests that it may be significantly easier to stabilize than traditional approaches based on waveguides or discrete elements. We also considered an experimental simplification where full dynamic control is not required and showed that, while not universal, with sufficient loops the unitary approximates a maximally mixing unitary. While we have specifically considered this architecture in the context of boson-sampling, the same scheme, or variations on it, may lend themselves to other linear optics applications, such as interferometry, metrology, or full-fledged LOQC.

\begin{acknowledgments}
This research was conducted by the Australian Research Council Centre of Excellence for Engineered Quantum Systems (Project number CE110001013). JPD would like to acknowledge the Air Force Office of Scientific Research.
\end{acknowledgments}

\bibliography{bibliography}

\appendix

\section{Alternate proof for universality} \label{app:alt_proof}

Here we will demonstrate an alternate approach that shows that the this scheme has the necessary ingredients to perform a full Reck \emph{et al.} decomposition and is thus universal. We present this for the case of \mbox{$n=3$} and then consider generalization.

\begin{figure}[!htb]
\includegraphics[width=0.8\columnwidth]{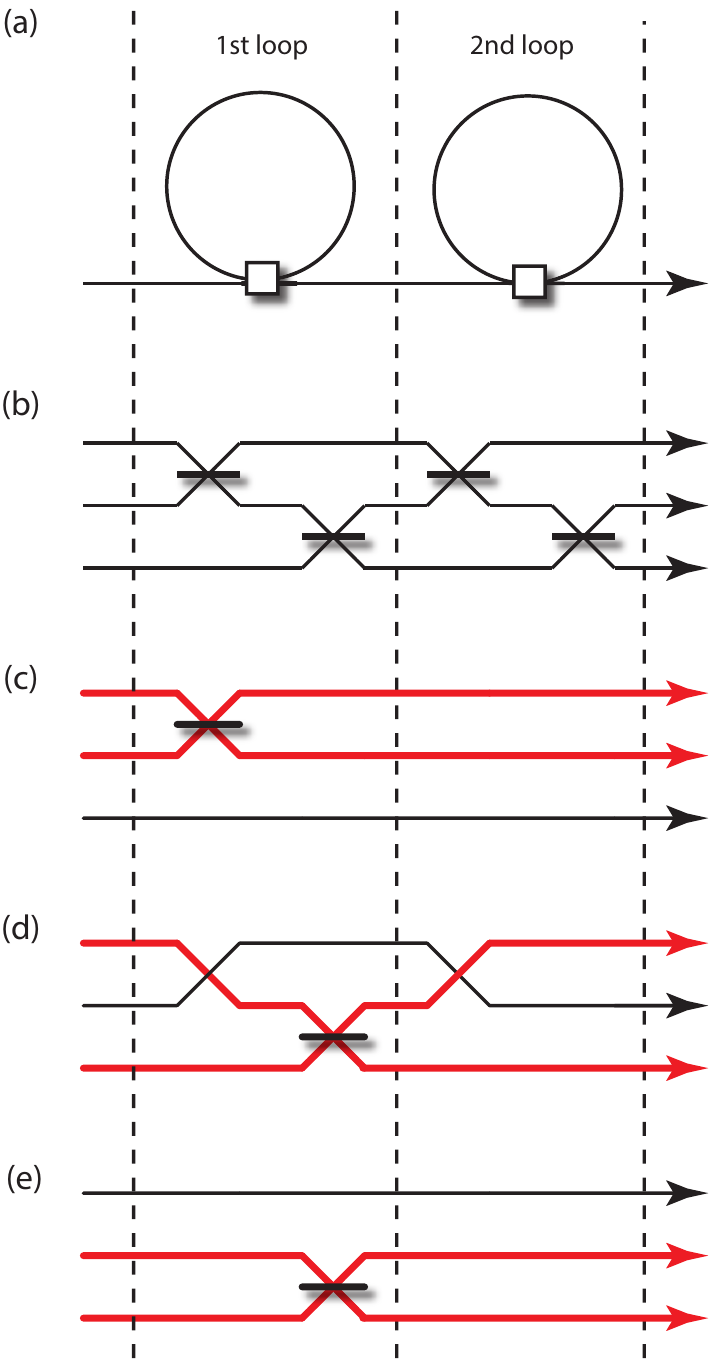} 
\caption{(Color online) (a) Two consecutive fiber loops in series. (b) The equivalent beamsplitter expansion for \mbox{$n=3$}. Each subsequent loop will add another diagonal array of beamsplitters. Setting the beamsplitter ratios, the two loops can implement an arbitrary beamsplitter between any pair of modes: (c) between modes 1 and 2, (d) between modes 1 and 3, (e) between modes 2 and 3. Thick red lines show the paths of the modes to which the beamsplitter operation is to be applied. In (c\--e) the beamsplitters not used to perform the interference are set as either completely reflective or transmissive in order to route the respective modes to the beamsplitter at which the interference is taking place.} \label{fig:universality}
\end{figure}

In Fig.~\ref{fig:universality} we show how two consecutive loops in series (a) produce an equivalent beamsplitter network (b), that allows for arbitrary pairwise beamsplitter interactions between any pair of two of the three modes (c-e). This is only possible with a dynamically controlled switch, which allows some of the beamsplitters to implement permutations, with one of the beamsplitters implementing the desired interaction. Having established that arbitrary pairwise beamsplitter interactions are possible, it follows that with subsequent iterations of more fiber loops, an arbitrary three mode network may be constructed.

\begin{figure}[!htb]
\includegraphics[width=0.8\columnwidth]{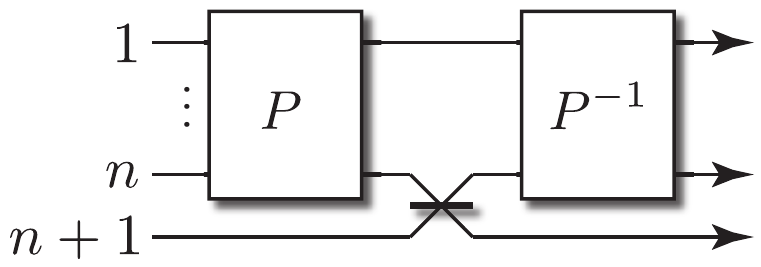} 
\caption{Generalizing the universality argument presented in Fig.~\ref{fig:universality} to arbitrary $n$. The argument is inductive. It follows from the universality of $n$ that an arbitrary $n$ mode unitary can be applied. We choose a permutation that the first of the desired modes be routed to the beamsplitter, which interacts with mode \mbox{$n+1$}. Then the inverse permutation is applied, leaving us with a network that implements an arbitrary beamsplitter operation between one of the first $n$ modes and the \mbox{$(n+1)$th} mode. It follows inductively that an arbitrary beamplsitter operation can be applied between any pair of modes for any $n$.} \label{fig:induction}
\end{figure}

Finally, we must generalize this universality result to the case of $n$ modes. In Fig.~\ref{fig:induction} we demonstrate that an inductive argument shows that if universality is possible for $n$ modes, it must also apply for \mbox{$n+1$} modes. Thus, universality applies for arbitrary $n$.

\end{document}